\documentclass[printer]{aa}
\usepackage{hyperref }
\usepackage{graphicx }
\usepackage{natbib   }
\usepackage{color    }
\usepackage{array    }
\usepackage{txfonts  }
\usepackage{breakurl }
\usepackage{multirow }
\usepackage{multicol }
\usepackage{booktabs }
\usepackage{dcolumn  }
\usepackage{latexsym }
\usepackage{subfigure}
\usepackage{amstext  }
\usepackage{amsmath  }
\usepackage{amssymb  }
\usepackage{psfrag   }
\usepackage{txfonts  }

\usepackage{ulem}

\def\hh{$^{\mathrm h}$}                  % hours
\def\mm{$^{\mathrm m}$}                  % minutes
\def\ss{$^{\mathrm s}$}                  % seconds
\def\d{$^\circ$}                         % degrees
\def\m{$^\prime$}                        % arcmin
\def\s{$^{\prime\prime}$}                % arcsec
\def\SB{{\itshape SB}}                   % surface brightness
\def\Chandra{{\itshape Chandra}}         % Chandra X-ray telescope's name
\def\XMM{{\itshape XMM-Newton}}          % XMM-Newton X-ray telescope's name
\def\HESSsmall{HESS~J1826$-$130}         % the 'small' HESS source
\def\HESSbig{HESS~J1825$-$137}           % the 'huge' HESS source
\begin{document} 

\title{A new study towards PSR~J1826$-$1334 and PSR~J1826$-$1256\\in the region of HESS~J1825$-$137 and HESS~J1826$-$130}

\titlerunning{A new study of PSR J1826$-$1334 and PSR J1826$-$1256}

\author{L. Duvidovich  \inst{1,2}
   \and E. Giacani     \inst{1,3}
   \and G. Castelletti \inst{1,2}
   \and A. Petriella   \inst{1,4}
   \and L. Supan       \inst{1,2}}

\authorrunning{Duvidovich et al.}

\institute{CONICET - Universidad de Buenos Aires, Instituto de Astronom\'{\i}a y F\'{\i}sica del Espacio (IAFE), Buenos Aires, Argentina.\\
           \email{duvidovich@iafe.uba.ar}
      \and Facultad de Ciencias Exactas y Naturales, Universidad de Buenos Aires, Buenos Aires, Argentina.
      \and Facultad de Arquitectura Dise\~no y Urbanismo, Universidad de Buenos Aires, Buenos Aires, Argentina.
      \and Ciclo Básico Común, Universidad de Buenos Aires, Buenos Aires, Argentina.}

\offprints{L. Duvidovich}

\date{Received 7 November 2018 / Accepted 21 January 2019}

\abstract{}{The goal of this paper is to detect synchrotron emission from the relic electrons of the crushed pulsar wind nebula (PWN) HESS~J1825$-$137 and to investigate the origin of the $\gamma$-ray emission from HESS~J1826$-$130.} 
{The study of HESS~J1825$-$137 was carried out on the basis of new radio observations centred at the position of PSR~J1826$-$1334 performed with the Karl G. Jansky Very Large Array at 1.4 GHz in configurations B and C. To investigate the nature of HESS~J1826$-$130, we reprocessed unpublished archival data obtained with {\XMM}.} 
{The new radio continuum image towards PSR~J1826$-$1334 reveals a bright radio source, with the pulsar located in its centre, which suggests that this feature could be the radio counterpart of the compact component of the PWN detected at high energy. The new 1.4~GHz radio data do not reveal emission with an extension comparable with that observed in $\gamma$-rays for the HESS~J1825$-$137 source. 
On the other hand, the {\XMM} study of the region including PSR~J1826$-$1256 reveals an elongated non-thermal X-ray emitting nebula with the pulsar located in the northern border and a tail towards the peak of the very high energy source. The spectrum is characterized by a power law with a photon index going from 1.6 around the pulsar to 2.7 in the borders of the nebula, a behaviour consistent with synchrotron cooling of electrons. From our X-ray analysis we propose that HESS~J1826$-$130 is likely produced by the PWN powered by PSR J1826$-$1256 via the inverse Compton mechanism.}{}


\maketitle
                     
\section{Introduction}
\label{1}
\begin{figure*}
\center
\vspace{1cm}
 \includegraphics[width=0.7\textwidth]{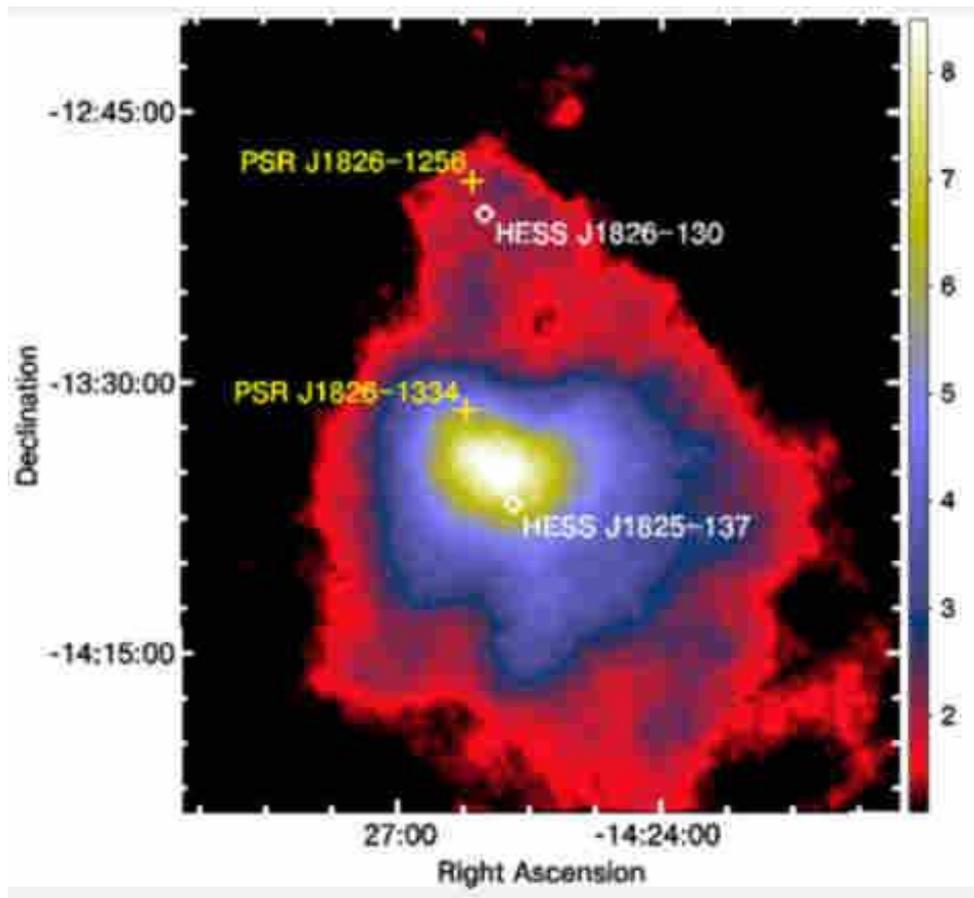}
\caption{HESS~J1825$-$137 and HESS~J1826$-$130 $\gamma$-ray excess map ($E$ $>$ 0.2~TeV) extracted from the on-line material of the HGPS \citep{HESS18a}.{\protect\footnotemark} The wedge is shown in units of $\mathbf{10^{-13}}$~ph~cm$^{-2}$~s$^{-1}$ the TeV flux. The plus signs indicate the position of the pulsars PSR~J1826$-$1334 and PSR~J1826$-$1256, and the diamonds the best-fit position of the TeV sources.}
\label{Figure1}
\end{figure*}

The H.E.S.S. Galactic Plane Survey (HGPS) has recently completed nine years of continuum observations in the 250$^\circ$-65$^\circ$ longitude range for latitudes $\rvert$\textit{b}$\rvert$ $\leq$ 3$^\circ$ \citep{HESS18a}. Observations towards the region containing the known pulsar wind nebula (PWN) HESS~J1825$-$137 show a new source named HESS~J1826$-$130, which emerges towards the north of HESS~J1825$-$137 and whose origin is still unknown (see Fig.~\ref{Figure1}).

HESS~J1825$-$137 is one of the brightest and most extensive PWN detected at very high energies (VHEs) \citep{HESS2018d}. It has been proposed that this TeV source is powered by the high spin-down luminosity ($\dot{E}$ = 2.8~$\times~10^{36}$ erg s$^{-1}$) pulsar \object{PSR~J1826$-$1334} \citep{Manchester05}. This pulsar is located at a distance of $\sim$4~kpc based on dispersion measurements \citep{Cordes02} and its spin-down age is $\sim$20~kyr. The pulsar is about 10\m~offset from the peak of HESS~J1825$-$137. This TeV source was the first to exhibit an energy dependent morphology characterized by a spectral steeping with increasing distance from the pulsar. In the GeV range, observations made with {\it Fermi}-LAT showed both morphological similarity and spectrum continuity with the H.E.S.S. data \citep{Grondin11}.

{\XMM} observations towards PSR~J1826$-$1334 show a bright and elongated core that is $\sim$30$^{\prime\prime}$ in size  with a hard photon index ($\Gamma$~$\sim$~$1.6$), which is embedded in a fainter diffuse component of emission that extends $\sim$$5^{\prime}$ with a softer photon index ($\Gamma$~$\sim$~$2.3$) and is located on the southern side of the pulsar \citep{Gaensler03}. This spectral behaviour is remarkably similar to that observed in $\gamma$-rays, except that the spatial scales are different (few arcmin in X-rays, fraction of a degree in $\gamma$-rays). Inside the compact nebula, {\Chandra} data reveal a small component ($\sim$7$^{\prime\prime}\times3^{\prime\prime}$) elongated in the NE-SW direction. The extended X-ray emission is detected in these observations up to at least $2^{\prime}.4$ south of the pulsar \citep{Pavlov08}. 

The offset and asymmetrical nature of the emission in the X-ray and $\gamma$-ray ranges have been explained by \citet{Gaensler03} and \citet{Aharonian2006a} using the scenario known as ``crushed PWN'' \citep{Blondin01}. In this framework, the supernova remnant (SNR) blastwave expands into an inhomogeneous interstellar medium (ISM). Consequently, an asymmetric reverse shock is generated, which returns from higher density regions first, reaching one side of the PWN sooner than the other side, then pushing the X-ray and TeV emission mainly towards regions of lower ambient density.

In the radio band, the region around PSR~J1826$-$1334 has been explored at 8.3~GHz by \citet{Frail97} and at 1.4~GHz by \citet{Gaensler00} using the Very Large Array (VLA). However, in both cases the observations failed to detect nebular emission down to the respective sensitivity limits (rms noise levels of 0.2 and 0.5~mJy~beam$^{-1}$ at 8.3 and 1.4~GHz, respectively).

\footnotetext{\url{https://www.mpi-hd.mpg.de/hfm/HESS/hgps/}. The image corresponds to the file {\itshape hgps\_map\_significance\_0.1deg\_v1.fits.gz}.}

Regarding HESS~J1826$-$130, it is a new unidentified extended TeV source previously hidden in the emission from the bright nearby PWN HESS~J1825$-$137 \citep{HESS18a}. Fitting the excess map with a 2D symmetric Gaussian, \citet{anguner17} estimated the best fit-position of HESS~J1826$-$130 to be R.A.=18\hh 26\mm 0\ss.2, Dec.=$-$13\d 02\m $1^{\prime\prime}.8$~(J2000) with an extension of {0.17\d} $\pm$ $0.02^\circ_\mathrm{stat} \pm 0.05^\circ_\mathrm{sys}$. 

Based on their spatial coincidence, HESS~J1826$-$130 has been associated with the \object{PWN~G18.5$-$0.4} (also known as `Eel' PWN), which is a diffuse and weak non-thermal X-ray nebula detected with {\Chandra} \citep{Roberts07}. This PWN is probably powered by the pulsar PSR~J1826$-$1256, which lies within the VHE emission region $5^{\prime}.4$ offset from its centroid, and now is notable for being one of the brightest radio-quiet $\gamma$-ray pulsars \citep{Acero15}. Nevertheless, within the extension of HESS~J1826$-$130 there are other potential sources of energetic particles capable to power the TeV emission, such as radio SNRs and HII regions \citep{paron13}, and this fact hampers a decisive conclusion on the origin of the $\gamma$-rays. 

In this paper we report on new high-resolution and high-sensitivity radio observations carried out with the Karl G. Jansky Very Large Array (JVLA)\footnote{The Very Large Array of the National Radio Astronomy Observatory is a facility of the National Science Foundation operated under cooperative agreement by Associated Universities, Inc.} towards PSR~J1826$-$1334 with the aim to detect the radio counterpart of the PWN detected at high and VHEs. In addition, we present a new X-ray study carried out with archival data acquired with the {\XMM} satellite of the PWN~G18.5$-$0.4 to investigate its connection with HESS~J1826$-$130 and shed light on the origin of this TeV source.

\section {Observations and data reduction}
\subsection {New radio observations}
We carried out observations centred on PSR~J1826$-$1334 using the B and C configurations of the JVLA.%
\footnote{Further information on the JVLA array configurations is reported at \url{https://science.nrao.edu/facilities/vla/docs/manuals/propvla/array_configs}.} 
The data were taken at L band using the wide-band 1 GHz receiver system centred at 1.4 GHz, which consists in 16 spectral windows with a bandwidth of 64 MHz each, spread into 64 channels. These observations are under the project 12A-166, which was intended to carry out deep full-synthesis imaging of intensity of the nebular radio emission from the vicinity of the scientific target source. 

The observing details are summarized in Table~\ref{data}. All the data were reduced using the capabilities within the CASA software infrastructure. 

\begin{table}[h!]
\centering
\caption{Summary of JVLA observations}
\label{data}
\begin{tabular}{ccc}\hline\hline
%-----------------------------------------------------------------
  Date           &   Config.  &  Integration time [min]  \\ \hline
%-----------------------------------------------------------------
  2012 ~Feb. 05  &      C     &         90               \\
  2012 ~Feb. 18  &      C     &         90               \\
  2012 ~Jun. 20  &      B     &         90               \\
  2012 ~Jun. 21  &      B     &         60               \\ \hline
%-----------------------------------------------------------------
\end{tabular}
\end{table}

After an initial data inspection and editing, the calibration was done according a standard procedure separately for each of the four observing dates. This includes determinations of antenna position and delay corrections, as well as complex antenna gains and bandpass solutions. For the four datasets observations of the primary flux density calibrator \object{3C~286} were used, in reference to a model of the flux distribution of the calibrator source, to calculate the antenna-based gain (amplitude and phase) corrections and to derive the flux of the phase calibrator \object{J1834$-$1237}. The source 3C~286 was also used to correct the bandpass. 

After the calibration was applied, the data of the scientific target PSR~J1826$-$1334 was split out from all of the observations listed in Table~\ref{data} and the individual files were imaged together using the capability incorporated in the CASA task tclean. Two rounds of phase-only self-calibration were performed to improve the dynamic range. It was found that a further amplitude calibration iteration was not necessary as it showed very little additional improvements on the image results. The self-cal process was made with a multi-scale technique with three different scale sizes and a robust parameter of 0.5. A wide-field imaging technique based on the W-projection algorithm was also applied during the imaging process to take account the non-coplanar nature of the JVLA baselines as a function of the distance from the phase centre. 

The synthesized beam of the final L-band image after combining the four datasets described in Table~\ref{data} and including primary beam corrections is 9.$^{\prime\prime}$24$\times$6.$^{\prime\prime}$43 with a position angle PA=0$^{\circ}$.05. The sensitivity achieved in the L-band image, without primary beam corrections applied, is 0.24~mJy~beam$^{-1}$.

\subsection {X-Ray data}
We reprocessed unpublished archival {\XMM} data (obs. ID \#0744420101) towards \object{PSR~J1826$-$1256} to analyse the properties of the extended X-ray emission. The EPIC-pn camera was operated in the small-window mode, which partially covers the extent of the PWN around PSR~J1826$-$1256. On the other hand, EPIC-MOS1 and MOS2 cameras were set in the full-frame mode and hence mapped the full extension of the nebula. We employed the two EPIC-MOS cameras for the analysis of the X-ray emission of the PWN and the three cameras for the spectral study of the pulsar. 
   
We reduced the observation and extracted the scientific products using the software packages SAS 16.1.0 and Heasoft 6.22.1, 
following standard procedures described in the SAS Sciences Threads.\footnote{\url{https://www.cosmos.esa.int/web/xmm-newton/sas-threads}.} After obtaining calibrated event files using the latest current calibration files (CCF), we constructed a light curve in the high-energy band ($> 10$~keV) to filter periods of high count-rate. We applied FLAG = 0 and PATTERN $\leq 12$ and 4 for MOS and pn cameras, respectively. 
We obtained filtered event files with 124, 128, and 91 ks for the MOS1, MOS2, and pn cameras, respectively. We constructed individual images from MOS1 and MOS2 cameras in the 1-7~keV band, the energy range where we detect emission, and combined these into a mosaicked and exposure-corrected image with a spatial scale of 3\s/pix. The final image was smoothed by convolving it with a 2D Gaussian function with $\sigma = 3$ pixel.

\section{Results and discussion}
\subsection{HESS~J1825$-$137}
\begin{figure*}
\center
\vspace{1cm}
\includegraphics[width=0.7\textwidth]{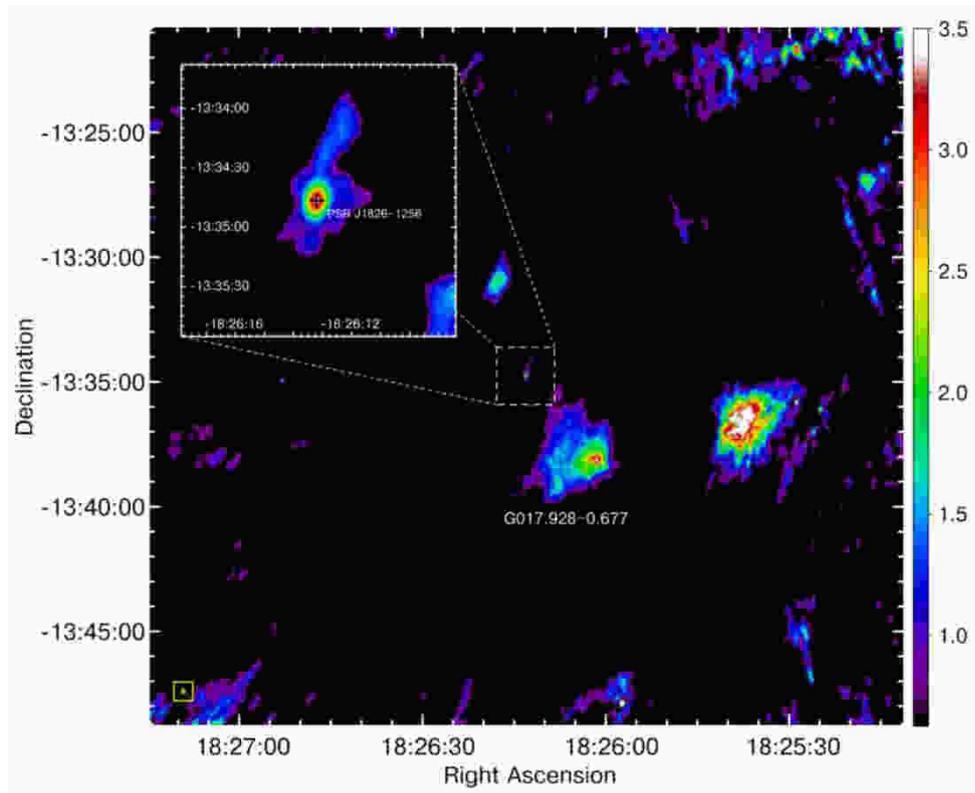}
\caption{Main panel shows the radio continuum image at 1.4 GHz of the region surrounding the pulsar PSR J1826$-$1334 constructed using B and C arrays of the JVLA. The intensity scale is based on a square root relation in units of mJy beam$^{-1}$. The final beam size, shown in the bottom left corner, is 9.$^{\prime\prime}$24$\times$6.$^{\prime\prime}$43 while the rms noise level is 0.24~mJy~beam$^{-1}$ without considering attenuation of the primary beam. The catalogued HII region G017.928$-$00.677 is also labelled \citep{anderson11}. The inset shows the radio continuum emission on the region surrounding the PSR J1826$-$1334, whose position is marked by a black plus sign.} 
\label{Figure2}
\end{figure*}

The main panel of Fig.~\ref{Figure2} shows the new JVLA radio image at 1.4 GHz centred on the position of PSR~J1826$-$1334. As the primary beam of the interferometer at the mentioned frequency is about 30$^{\prime}$, the data only cover the northern half of the source HESS~J1825$-$137. Especially prominent is the emission from two objects in the surveyed field. One of these sources, with its brightness peak at R.A.=18\hh 26\mm 1\ss.7, Dec.=$-$13\d 38\m 14$^{\prime\prime}$.6~(J2000),  is in positional coincidence with the HII region G017.928$-$0.677 and most likely represents thermal emission associated with ionized gas \citep{anderson11}. The catalogued kinematical distance to this thermal source is $\sim$13~kpc, based on recombination-line emission detected at $\sim$39~km~s$^{-1}$ with the NRAO Green Bank 140-foot telescope \citetext{\citealt{anderson11}, see also \citealt{roman-duval09}}. The emission centred at R.A.=18\hh 25\mm 37\ss.5, Dec.=$-$13\d 36\m 41$^{\prime\prime}$~(J2000) is catalogued as NVSS~ J182538$-$133637 in the 1.4~GHz NRAO VLA Sky Survey \citep{condon98} and referred to as source ``C'' in \citet{Pavlov08}. To the best of our knowledge, the literature does not contain a counterpart that helps to discern its nature.

A zoomed view of the region surrounding  PSR~J1826$-$1334 is shown in the upper left corner of Fig~\ref{Figure2}. This image shows, for the first time, emission from a bright region  around the pulsar with a size of about 20\s, in good agreement with the size of the X-ray compact component of the PWN reported by \citet{Gaensler03}. We estimated the total flux density of this structure, detected at 4$\sigma$ above the noise level, in $\sim$4~mJy. The brightest central portion of the compact radio component, which is comparable in size to the angular resolution of our observations,  harbours the pulsar. Its contribution to the flux calculated over all the compact component is about 3~mJy. We highlight that this result is largely consistent with the flux density determination for PSR~J1826$-$1334 reported by \citet{kijak07}, on the basis of the total on-pulse energy measured over the pulse period around 1~GHz using the Giant Metrewave Radio Telescope \citetext{see also the compatible estimates presented in \citet{jankowski18} and references therein}.

The enlargement included in the new image at 1.4~GHz (Fig.~\ref{Figure2}) also shows, at 2$\sigma$ above the noise, a feature extending to the north of the pulsar position. Even though the detection occurs at a low level, the internal structure of this faint feature suggests that it could be real. If it were, we are likely observing part of the X-ray PWN. To confirm this, further observations with higher sensitivity are required. In addition, further polarization and spectrum studies are needed since a high degree of polarization ($\gtrsim 10\% $) and a flat radio spectral index ($ -0.3 \leq \alpha \leq 0$, S $\propto \nu^{\alpha}$,  where $S$ is the flux density, $\nu$ the frequency and $\alpha$ the radio spectral index) are two unmistakable properties of PWNe in the radio band \citep{Gaensler06}.

The non-detection in the radio band of a PWN detected at high energies is not an unusual result. Indeed, there are a large number of pulsars for which the search of radio PWNe yielded negative results \citep [e.g.][]{Gaensler00,Giacani14,cas2016}. In most of these cases, the failure to detect the nebular emission has been explained as a consequence of  either  a high magnetic field of the pulsar that inhibits the production of synchrotron radiation at longer wavelengths, or  severe adiabatic losses that occur in young and energetic pulsars producing under luminous radio PWNe if they evolve in a very low ambient density ($\sim$0.003~cm$^{-3}$). 
Nevertheless, neither of the two conditions are satisfied for PSR~J1826$-$1334. On one hand, taking into account the parameters of the pulsar ($P$ = 101~ms and $\dot{P}$ = 7.52 $\times$ 10$^{-14}$~s~s$^{-1}$, \citet{Clifton98}), we estimated its magnetic field strength in about 2.8~$\times$~10$^{12}$~G, a value typical of young pulsars in SNRs. On the other side, the requirement of an ambient of very low density is not consistent with the presence of several dense molecular regions in the vicinity of the pulsar at its kinematic distance \citep{voisin16}.

Even though the two conditions invoked to account for the lack of an associated radio nebula fail to explain the case for PSR~J1826$-$1334, we can quantify our non-detection. To do this, we characterized the luminosity integrated between 10$^{7}$~Hz and 10$^{11}$~Hz of the putative radio PWN by L$_{\mathrm{R}} \equiv \epsilon\,\dot{\rm E}$~erg~s$^{-1}$, where $\dot{\rm E}$ is the associated spin-down luminosity of the pulsar and $\epsilon$ is the efficiency of conversion of $\dot{\rm E}$ in radio emission. Assuming that typical values for the efficiency ($\epsilon$$\simeq$10$^{-4}$) and the spectral index ($\alpha$=$-$0.3) measured in radio PWNe \citep{Gaensler06} are also valid for the case of PSR~J1826$-$1334, we obtained an upper limit for the flux density at 1.4~GHz of $\sim$400~mJy; this upper limit is calculated using the relation $S= 2.15 \times 10^{5} \epsilon \dot{\rm E}_{34}$ $d^{-2}$~mJy, where $d$ is the distance to the pulsar in kpc and $\dot{\rm E}_{34}$ is in units of 10$^{34}$~erg. If the extension of the supposed PWN were at least the same as that of the X-ray nebula observed with {\XMM} satellite (ellipsoid $\sim$5$^\prime$ $\times$ 4$^\prime$ in size), then the limit of the detection would be $\sim$0.3~mJy~beam$^{-1}$, i.e. of the order of the rms noise level of our radio observations. Morever, in the case in which the size of the radio PWN were comparable to that observed in $\gamma$ rays, the corresponding limit for its detection in radio would be even one order of magnitude lower.

\subsection{HESS~J1826$-$130}

\subsubsection{X-ray imaging and spectral analysis}
\label{X_spec}
\begin{figure*}[ht!]
\centering
\includegraphics[width=0.90\linewidth]{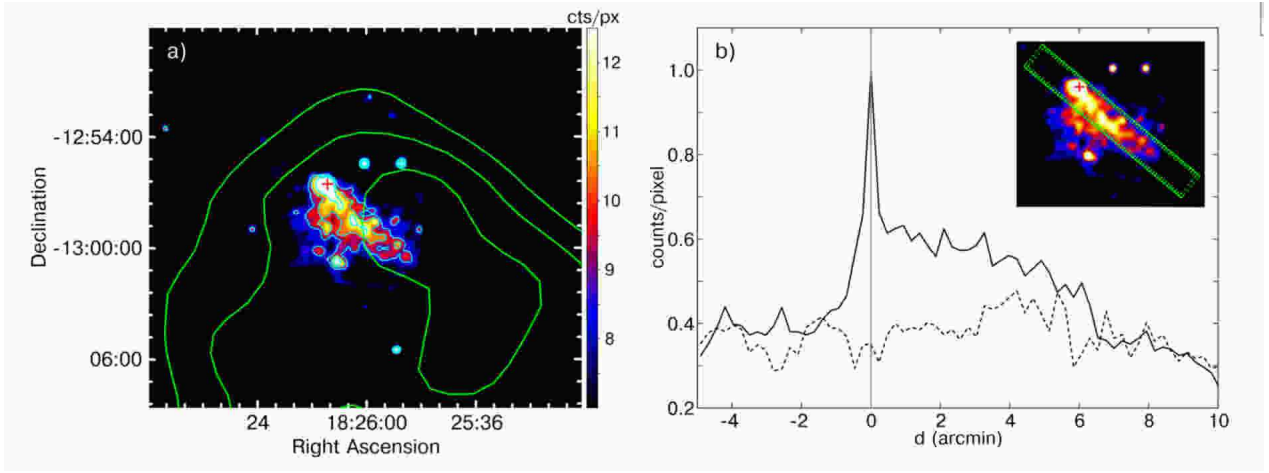}
\caption{\bf a. \rm  X-ray emission towards PSR~J1826$-$1256 in the 1-7~keV energy band. Cyan contours correspond to 9.5, 10.5, and 12 counts/pixel. The position of the pulsar is indicated with a plus sign. Green contours trace the TeV emission from HESS~J1826$-$130 at levels of (1.5, 1.8, and 2.1)~$\times$~$10^{-13}$~ph~cm$^{-2}$~s$^{-1}$. \bf b. \rm A brightness profile along the PWN (solid line) and the adjacent background (dashed line). The rectangular box in the inset image  indicates the regions from which the radial profile was extracted.}
\label{Figure3}
\end{figure*}

Figure~\ref{Figure3}a shows the X-ray emission towards PSR~J1826$-$1256 in the 1-7~keV energy band with overlaid contours of the TeV emission from HESS~J1826$-$130, for reference, while panel b of Fig.~\ref{Figure3} indicates a brightness profile. This profile was constructed from a mosaicked image using data from the MOS1 and MOS2 cameras with a spatial scale of $1.5^{\prime\prime}/\rm{pix}$. No exposure correction and smoothing was applied. 
We defined rectangular boxes of $130^{\prime\prime}\times 14^{\prime\prime}$  placed along the most intense X-ray emission. For the background, we used rectangular regions of $60^{\prime\prime}\times 14^{\prime\prime}$ orientated in the same way as the source boxes and placed outside the diffuse emission. We defined the box containing the pulsar J1826$-$1256 as the reference position to measure boxes distances. 

\begin{figure*}[ht!]
\centering
\includegraphics[width=0.9\linewidth]{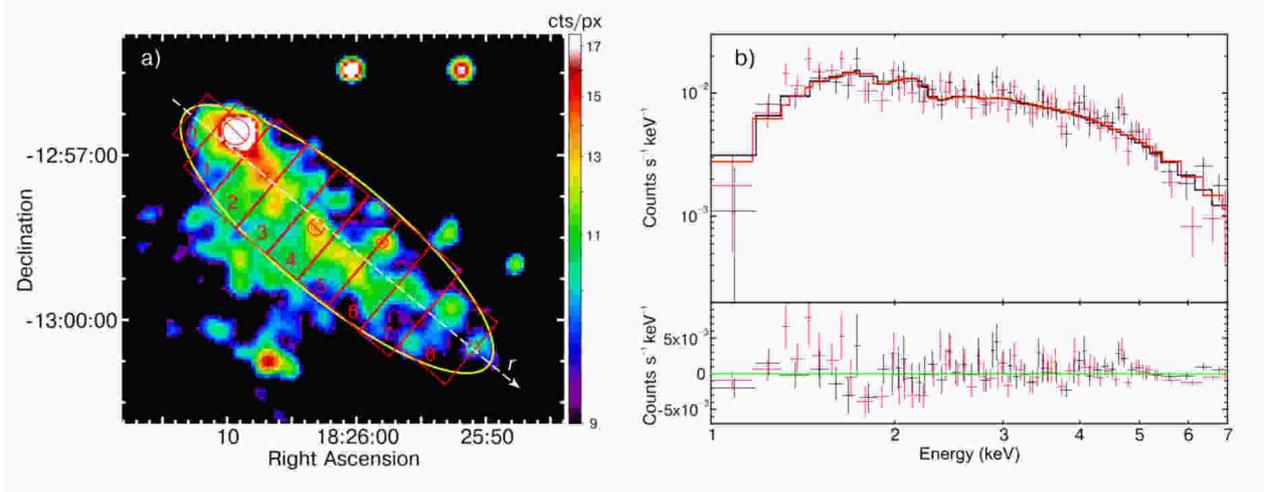}
\caption{\bf a. \rm Regions used for the spatially resolved spectral analysis. Red circles are likely point sources excluded from the extraction regions. The circle inside box 1 was used to exclude the emission of the pulsar from the surrounding diffuse emission. \bf b. \rm MOS1 (black) and MOS2 (red) spectra of the bar in the 1-7~keV band. The solid lines represent the best fit for an absorbed power-law model.  }
\label{Figure4}
\end{figure*}

The large-exposure {\XMM} observation allows us to identify some morphological features not detected in previous imaging studies performed with {\Chandra}. Indeed, the asymmetric and non-uniform distribution of the X-ray emitting gas around PSR~J1826$-$1256 is shown in Fig.~\ref{Figure3}. Bright emission is observed ahead of the pulsar up to a distance of $\sim$$1^{\prime}$ (Fig.~\ref{Figure3}b). We notice, however, that this emission may be due to projection effects. An enlargement showing the detailed X-ray morphology is presented in Fig.~\ref{Figure4}a. 
The bulk of the emission comes from an elongated feature with an elliptical shape of $\sim$$6^{\prime}\times2^{\prime}$ (major $\times$ minor-axis), which appears brighter around the pulsar and extends towards the southwest in the direction of the TeV source. We refer to this feature as the ``bar'' and we associate it with the ``tail'' identified by \citet{Roberts07} in the {\Chandra} image. These authors suggested that the shape of the X-ray PWN could be caused by the motion of the pulsar which leaves behind a high-energy trailing nebula. The bar is surrounded by faint and diffuse emission that is more prominent towards the southeast. PSR~J1826$-$1256 appears slightly displaced from the apparent axis of the bar (indicated with a green line in Fig.\ref{Figure4}a).

Several point-source candidates are seen in the field in the {\Chandra} Source Catalog (CSC; \citealt{eva10}) and in the {\XMM} Serendipitous Source Catalog (3XMM DR8; \citealt{Rosen16}). We identified the brightest sources that appear over the bar (indicated with red circles in Fig.\ref{Figure4}a) and excluded these from the spectral analysis. To obtain the overall properties of the PWN, we extracted the spectrum of the bar from the ellipse shown in Fig. \ref{Figure4}a. The background was chosen from a circular region free of diffuse emission and excluding the point sources over it. We obtained individual spectra for MOS1 and MOS2 with minimum of 30 cts/bin. Inspection and fitting of the spectra was done using Xspec 12.9.1. The spectra are hard and show a continuum with no hint of emission lines (see Fig.~\ref{Figure4}b). 
We simultaneously fitted the MOS1 and MOS2 spectra in the 1-7~keV energy band with an absorbed power-law model ($wabs \times powerlaw$). The results are shown in Table \ref{tab_X1}. The reported flux was calculated in the 0.5-8.0~keV energy band and this flux is\ used in Sect.~\ref{hess} to analyse the possible connection between the X-ray PWN and HESS~J1826$-$130.   

\begin{table*}[ht!]
\caption{Spectral analysis parameters of the diffuse emission from the PWN and the pulsar PSR~J1826$-$1256. For the PWN we used MOS1 and MOS2 cameras, while data from the pn camera was also considered for analysing the emisson from the pulsar. Spectral fitting with an absorbed power law was performed in the 1-7~keV energy band. The symbol ($^{*}$) indicates that the parameter was fixed during the fitting process. The value $F$ is the absorption corrected flux calculated in the 0.5-8.0~keV energy band. Quoted errors correspond to the 90\% confidence range.}
\label{tab_X1}
\small
\centering
\setlength{\extrarowheight}{5pt}
\begin{tabular}{cccccc}
\hline\hline
%-------------------------------------------------------------------------------------------------------------------------------------------------
Region        & Total counts & $\chi^{2}_{\nu}$ (d.o.f.) & $N_\mathrm{H}$        & $\Gamma_X$       & $F$~(0.5-8.0~$\rm{keV})$            \\
              &              &                           & [$10^{22}$ cm$^{-2}$] &                  & [$10^{-13}$~erg~s$^{-1}$ cm$^{-2}$] \\\hline
%-------------------------------------------------------------------------------------------------------------------------------------------------
Bar (ellipse) & 24600        & 1.02 (212)                & $1.89 \pm 0.26$       & $1.79 \pm 0.17$  & $13.4 \pm 0.4$                      \\
Pulsar        & 3300         & 1.05 (156)                & $1.89^*$              & $1.42 \pm 0.12 $ & $1.22 \pm 0.06$                     \\       
Pulsar        & 3300         & 1.04 (155)                & $1.59 \pm 0.42$       & $1.26 \pm 0.25 $ & $1.14 \pm 0.06$                     \\\hline
%-------------------------------------------------------------------------------------------------------------------------------------------------
\end{tabular}
\end{table*}

\begin{figure}[h!]
\centering
\includegraphics[width=0.85\linewidth]{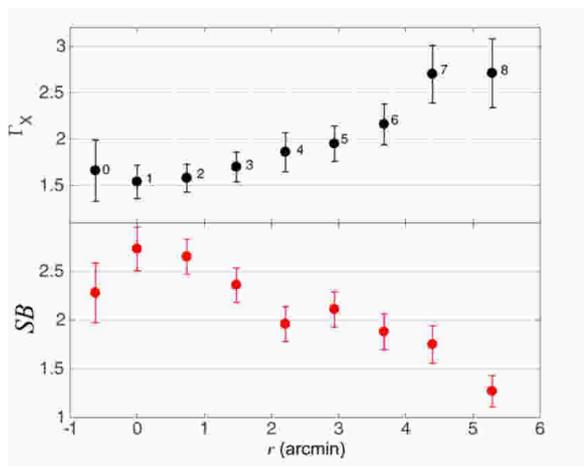}
\caption{Variation of the X-ray photon index $\Gamma_X$ (top) and the surface brightness {\SB} (bottom) with respect to the distance $r$ to box 1 (labelled in Fig.~\ref{Figure4}a). The {\SB} is expressed in units of $10^{-17}$~erg~s$^{-1}$~cm$^{-2}$~arcsec$^{-2}$.}
\label{Figure5}
\end{figure}

Taking advantage of the high number of counts of the {\XMM} observations, we performed a spatially resolved spectral analysis to look for variations of the photon index $\Gamma_X$ along the diffuse emission. We extracted individual spectra from nine rectangular boxes located along the bar (see Fig.~\ref{Figure4}a) from MOS1 and MOS2 cameras excluding the point sources. We defined the distance $r$ among boxes as the separation between their centres and the centre of box 1, which contains the pulsar J1826$-$1256. This is indicated in Fig.~\ref{Figure4}a with a green arrow. Spectra were binned to a minimum of 30 cts/bin. For each box, the two spectra are simultaneously fitted with a power-law model in the 1-7~keV energy band, keeping the hydrogen column density frozen to the best-fit value obtained for the whole diffuse emission, namely $N_\mathrm{H}=1.89\times10^{22}$~cm$^{-2}$. 
The results are summarized in Table~\ref{tab_X2} and plotted in Fig.~\ref{Figure5}. The top panel in Fig.~\ref{Figure5} shows the variation of the photon index variation $\Gamma_X$ with respect to $r$. The softening of the spectrum is remarkable, where $\Gamma_X$ increases from $\sim$1.6 around the pulsar (boxes 0-2) up to $\sim$ 2.7 in the outer regions (boxes 7 and 8). This behaviour has been observed in several Galactic PWNe, such as G0.9+0.1 \citep{porquet03}, 3C~58 \citep{bocchino01a}, and G21.5$-$0.9 \citep{safi01}, among others. The spectral softening is a consequence of the underlying synchrotron radiation mechanism responsible for the non-thermal emission in the keV band. Highly energetic electrons cool faster than low energetic electrons after having travelled a short distance from the powering pulsar while they diffuse throughout the pulsar wind. 
On the other hand, the surface brightness {\SB}, which is defined as the ratio of the absorption corrected flux in the 1-7~keV band and the area of the corresponding region, decreases with increasing distance to the powering pulsar, as shown  in the bottom panel of Fig.~\ref{Figure5}. This is a consequence of the synchrotron radiation mechanism responsible for the X-ray emission from PWNe \citep{holler12}.   

For the spectral study of the pulsar PSR~J1826$-$1256 we used the three cameras (MOS1, MOS2, and pn) and extracted the corresponding spectra from a circular region of radius $14^{\prime\prime}$. Spectra were binned to a minimum of 15 cts/bin. For the background, we defined an annular region with inner and outer radii of $20^{\prime\prime}$ and $80^{\prime\prime}$, respectively. X-ray emission from pulsars may correspond to black-body (BB) emission from the neutron star surface and/or to non-thermal synchrotron emission from the magnetosphere. To look for non-thermal emission from the pulsar, we fitted the spectra with an absorbed power-law model. Letting $N_\mathrm{H}$ to vary freely or freezing it to the best-fit value of the diffuse emission ($N_\mathrm{H} = 1.89\times 10^{22}$ cm$^{-2}$) gives statistically acceptable fits ($\chi^{2}_{\nu} \simeq 1.05$) and model parameters that agree within the uncertainties (see Table \ref{tab_X1}). 
Fitting the spectra with an absorbed BB ($wabs \times blackbody$) and $N_\mathrm{H}$ fixed to $1.89\times 10^{22}$ cm$^{-2}$ yields a statistically poorer fit ($\chi^{2}_{\nu} \simeq 1.45$), which fails to model the low-energy part of the spectra ($E<3$~keV). Letting $N_\mathrm{H}$ to vary freely results in a better fit ($\chi^{2}_{\nu} \simeq 1.03$) but a hydrogen column density of $\sim$0.2$\times 10^{22}$ cm$^{-2}$ is obtained in this case, which is about an order of magnitude lower than the column density for the diffuse emission. If PSR~J1826$-$1256 is powering the PWN, we would expect a similar ISM absorption. Moreover, the obtained BB temperature ($\sim$1.6~keV) is higher than expected for young pulsars, which usually have BB temperatures $\lesssim$$\,0.5$~keV \citep{mereghetti11}. We therefore conclude that the model which better fits the X-ray emission from PSR~J1826$-$1256 is a power law with the hydrogen column density similar to that obtained towards the PWN, as expected if they are associated sources. 

\begin{table*}[ht!]
\caption{Parameters of spatially resolved spectral analysis of the bar feature. We used MOS1 and MOS2 cameras. The value $r$ is the distance between the centre of each box and box 1 (see Fig.~\ref{Figure4}a), where PSR~J1826$-$1256 appears projected. Spectral fitting with an absorbed power law was performed in the 1-7~keV energy band, keeping $N_\mathrm{H}$ frozen to $1.89 \times 10^{22}$~cm$^{-2}$. Quoted errors indicate the 90\% confidence level uncertainties.}
\label{tab_X2}
\small
\centering
\setlength{\extrarowheight}{5pt}
\begin{tabular}{ccccccc}
\hline\hline
%---------------------------------------------------------------------------------------------------------------------------------------------------
Box &  $r$      & area           & Counts & $\chi^{2}_{\nu}$ (d.o.f.) & $\Gamma_X$       & {\SB}~(1-7~$\rm{keV})$                           \\
    &  [arcmin] & [arcsec$^{2}$] &        &                           &                  & [$10^{-17}$~erg~s$^{-1}$~cm$^{-2}$~arcsec$^{-2}$]\\\hline
%---------------------------------------------------------------------------------------------------------------------------------------------------
0   &  $-0.62$  & 1650           & 1242   & 1.04 (39)                 & $1.66 \pm 0.34$  & $2.28 \pm 0.31$                                  \\
1   &  $0$      & 3160           & 2646   & 1.04 (84)                 & $1.54 \pm 0.18$  & $2.73 \pm 0.22$                                  \\
2   &  $0.74$   & 5100           & 4128   & 1.15 (129)                & $1.58 \pm 0.15$  & $2.65 \pm 0.18$                                  \\
3   &  $1.48$   & 5100           & 3967   & 0.96 (124)                & $1.70 \pm 0.16$  & $2.36 \pm 0.17$                                  \\
4   &  $2.21$   & 4833           & 3602   & 1.05 (111)                & $1.86 \pm 0.21$  & $1.96 \pm 0.18$                                  \\
5   &  $2.93$   & 5080           & 3782   & 0.90 (116)                & $1.95 \pm 0.19$  & $2.11 \pm 0.18$                                  \\
6   &  $3.66$   & 5080           & 3616   & 1.07 (113)                & $2.16 \pm 0.22$  & $1.88 \pm 0.18$                                  \\
7   &  $4.39$   & 5100           & 3501   & 0.87 (168)                & $2.70 \pm 0.31$  & $1.75 \pm 0.19$                                  \\
8   &  $5.27$   & 5534           & 2739   & 0.99 (86)                 & $2.71 \pm 0.37$  & $1.27 \pm 0.16$                                  \\\hline
%---------------------------------------------------------------------------------------------------------------------------------------------------
\end{tabular}
\end{table*}

\begin{table*}[h]
\caption{Parameters of PSR~J1826$-$1256, the X-ray PWN and HESS~J1826$-$130 for distances of 4.6 and 11.4~kpc. References are (1) this work, (2) \citet{abdo10}, and (3) \citet{HESS2018d}. Some of the parameters referred to this work were derived using parameters of Ref. 2 and/or 3.}
\label{table_disc}
\small
\centering
\setlength{\extrarowheight}{5pt}
\begin{tabular}{lccll}
\hline
%------------------------------------------------------------------------------------------------------------------------------------------------------------------------
 Parameter                                      & $D = 4.6$~kpc         & $D=11.4$~kpc          &                                                       & Ref.  \\\hline
%------------------------------------------------------------------------------------------------------------------------------------------------------------------------
 $\dot{E}$ [erg~s$^{-1}$]                       & \multicolumn{2}{c}{$3.60\times 10^{36}$}      & Pulsar's spin-down power                              & (2)   \\
 $\tau_c$ [yr]                                  & \multicolumn{2}{c}{$1.4\times10^4$}           & Pulsar's characteristic age                           & (2)   \\
 $\Gamma_{\mathrm{X}}^{\mathrm{PSR}}$           & \multicolumn{2}{c}{$1.42$}                    & Photon index of the pulsar in the keV band            & (1)   \\
 $L_{\mathrm{X}}^{\mathrm{PSR}}$ [erg s$^{-1}$] & $2.3 \times 10^{32}$  &  $1.7 \times 10^{33}$ & Pulsar luminosity in the 0.5-8.0~keV band             & (1)   \\\hline
%------------------------------------------------------------------------------------------------------------------------------------------------------------------------
 $\Gamma_{\mathrm{X}}$                          & \multicolumn{2}{c}{$1.79$}                    & Photon index of the PWN in the keV band               & (1)   \\
 $L_{\mathrm{X}}$ [erg s$^{-1}$]                & $3.4 \times 10^{33}$  &  $2.0 \times 10^{34}$ & PWN luminosity in the 0.5-8.0~keV band                & (1)   \\
 $\eta_{\mathrm{X}}$                            & $9.4 \times 10^{-4}$  &  $5.8 \times 10^{-3}$ & Pulsar's efficiency in the keV band                   & (1)   \\\hline
%------------------------------------------------------------------------------------------------------------------------------------------------------------------------
 $\Gamma_{\mathrm{TeV}}$                        & \multicolumn{2}{c}{$2.04$}                    & TeV spectral index                                              & (3)   \\
 $L_{\mathrm{TeV}}$ [erg s$^{-1}$]              & $1.0 \times 10^{34}$  &  $6.5 \times 10^{34}$ & Luminosity of HESS~J1826$-$130 in the 1-10 TeV band   & (3)   \\
 $\eta_{\mathrm{TeV}}$                          & $2.9 \times 10^{-3}$  &  $1.8 \times 10^{-2}$ & Pulsar's efficiency in the TeV band                   & (1,3) \\\hline
%------------------------------------------------------------------------------------------------------------------------------------------------------------------------
\end{tabular}
\end{table*}

\subsubsection{Origin of TeV emission}
\label{hess}
In a search for counterparts to HESS~J1826$-$130, the X-ray PWN driven by the PSR~J1826$-$1256 appears to be the most plausible candidate. In a leptonic scenario the TeV emission is expected to arise from inverse Compton (IC) scattering between the ambient low energetic photons and the same population of electron producing synchrotron radiation in the keV band. In our analysis of the spectral behaviour (see Sect.~\ref{X_spec}), we found that the photon index $\Gamma_\mathrm{X}$ of the extended X-ray PWN is $1.8 \pm 0.2$, close to the value $\Gamma_{\mathrm{TeV}}=2.0 \pm 0.1$ of the TeV spectrum of HESS~J1826$-$130 \citep{HESS18a}. This fact suggests a scenario in which the X-ray and TeV emissions arise from the same population of electrons \citep{pav08}. Based on our X-ray results and assuming that the softening of the spectrum with the distance from the pulsar is due to cooling effects, we estimated the magnetic field $B$ in the nebula from the relation $\tau_{\mathrm{sync}}\sim 38B_{\mu\rm{G}}^{-3/2} E^{-1/2}_{\mathrm{keV}}$~kyr, where $B_{\mu\rm{G}}$ is the magnetic field expressed in $\mu \rm{G}$ and $E_{\mathrm{keV}} = E_{\mathrm{sync}}/ (1~\rm{keV})$ being $E_{\mathrm{sync}}$ the energy of photons produced in the synchrotron radiation. As a consequence of the radiative cooling, the synchrotron cooling time $\tau_{\mathrm{sync}}$ should be smaller than the age of the PWN, which we roughly  approximated to the pulsar's characteristic age $\tau_c \sim 14$~kyr. The condition $\tau_{\mathrm{sync}} < \tau_\mathrm{c}$, gives $B > 2$ $\mu\rm{G}$ for $E_{\mathrm{sync}} = 1.5$~keV (the peak of the X-ray spectrum shown in Fig.~\ref{Figure4}a). The corresponding energy of the electrons is $E_\mathrm{e} \sim 160E_{\mathrm{keV}}^{1/2}B_{\mu\rm{G}}^{-1/2}$~TeV. For $E_{\mathrm{sync}}=1.5$~keV and $B > 2$~$\mu$G, we constrained the energy of the synchrotron emitting electrons to be $E_\mathrm{e} < 150$~TeV.  

The population of relativistic electrons analysed in the paragraph above, with energy $E_\mathrm{e}$, can also interact with background photons of energy $\epsilon$ via IC scattering producing $\gamma$-ray photons in the TeV band, revealed as the VHE counterpart to the X-ray PWN. Hence, we estimated the energy of the TeV photons through the relation $E_{\mathrm{IC}} \sim 4 (E_\mathrm{e}/\rm{TeV})^2 \epsilon_{\mathrm{eV}}$~TeV, where $\epsilon_{\mathrm{eV}} = \epsilon / (1~\rm{eV})$. If the cosmic microwave background is the main source of background photons with $T \sim 3$~K, then $\epsilon \sim 3 \times 10^{-4}$~eV. Using $E_\mathrm{e} < 150$~TeV, we found that the energy of the photons produced by the IC scattering is $E_{\mathrm{IC}} < 30$~TeV. This analysis shows that the VHE counterpart to the X-ray nebula around PSR J1826$-$1256 is expected to radiate IC photons with energies up to a few tens of TeV. This result is largely compatible with the detection of $\gamma$ rays in the 0.5-40~TeV range from  HESS~J1826$-$130 \citep{anguner17}.

Another approach to analyse  the possible connection between PSR~J1826$-$1256, its PWN, and HESS~J1826$-$130, is to compare the high-energy properties of the system with those observed in others PWNe. To do this, we used the luminosities $L_{\mathrm{X}}$ and $L_{\mathrm{TeV}}$  to estimate the efficiency of pulsar's energy conversion in the keV and TeV bands as $\eta_{\mathrm{X}} = L_{\mathrm{X}} / \dot{E}$ and $\eta_{\mathrm{TeV}} = L_{\mathrm{TeV}} / \dot{E}$, where  $L_{\mathrm{X}}$ and $L_{\mathrm{TeV}}$ are the luminosities of the PWN in the 0.5-8.0 keV and 1-10 TeV energy bands, respectively. 
From Table~\ref{tab_X1}, the flux of the PWN in the 0.5-8.0~keV band is $F_{X}\sim 13.4\times 10^{-13}$~erg~s$^{-1}$~cm$^{-2}$. This translates into a luminosity $L_{\mathrm{X}}\sim 1.60\times 10^{32} D_{\mathrm{kpc}}^{2}$~erg~s$^{-1}$, where $D_{\mathrm{kpc}}$ is the distance in kpc. Using the $\gamma$-ray flux reported by \citet{anguner17}, the luminosity in the 1-10 TeV band is $L_{\mathrm{TeV}}\sim 5.01\times 10^{32} D_{\mathrm{kpc}}^{2}$~erg~s$^{-1}$. From the obtained luminosities, together with the pulsar spin-down power $\dot{E}=3.60\times 10^{36}$ erg s$^{-1}$, we calculated the efficiency of pulsar's energy conversion $\eta_{\mathrm{keV}} \sim 4.5 \times 10^{-5} D_{\mathrm{kpc}}^{2}$ and $\eta_{\mathrm{TeV}} \sim 1.4\times 10 ^{-4} D_{\mathrm{kpc}}^{2}$ as a function of the distance. 
In the case of PSR~J1826$-$1256 its distance is not well known. \citet{Wang11} suggested a distance to the pulsar of about 1.2~kpc, based on the correlation found between the efficiency in the GeV band and other pulsar parameters in a sample of $\gamma$-ray pulsars detected by {\it Fermi}-LAT. However, from a molecular study of a wide region containing the source HESS~J1826$-$ 130, \citet{voisin16} reported the presence of molecular gas in spatial coincidence with it at kinematical velocities between 60 and 80~km~s$^{-1}$ that correspond to intermediate near/far distances of about 4.6/11.4~kpc. Besides, the authors pointed out that a distance of 1.2~kpc to the pulsar implies a TeV $\gamma$-ray efficiency lower than the typical values. Thus, by adopting in our calculations the distances reported by \citet{voisin16} for the PSR-PWN-HESS source system, we compared their properties  with the population of Galactic PWNe and listed them in Table~\ref{table_disc}. Photon indexes $\Gamma$, in both keV and TeV bands, are in good agreement with those measured for other PWNe, and the obtained efficiencies $\eta_\mathrm{X}$ and $\eta_{\mathrm{TeV}}$ show that PSR~J1826$-$1256 has a spin-down energy sufficient to power the observed nebulae in X- and $\gamma$-rays \citep{kargal13, HESS2018d}.

We also mention that a number of HII regions, OB stars, and high-mass star-forming regions are located in the vicinity of the TeV source at a distance of $\sim$4~kpc \citep{paron13}. These objects are able to accelerate particles to multi-TeV energy via hadronic interactions with dense material \citep[e.g,][]{araudo08,br10}. At the distance of 4~kpc, \citet{voisin16} reported the presence of molecular gas overlapping the TeV source, thus the contribution to the $\gamma$-ray emission via a hadronic process cannot be completely ruled out. 
 
\section{Summary}
In this paper we presented the highest angular resolution and sensitivity radio continuum emission image from JVLA observatory at 1.4 GHz towards the pulsar PSR~J1826$-$1334. We sought to search for the radio counterpart to the PWN HESS~J825$-$137, which has been also detected in the X-ray band. The new radio data allowed us to detect an extended bright radio source with the pulsar located in its centre, in positional coincidence, and with a size similar to that of the compact component of the X-ray PWN. Although with the available radio data it is not possible to go deeper in the analysis of the origin of the observed emission, we suggest that it could be the radio counterpart of the PWN detected in the X- and $\gamma$-ray spectral regimes. Regarding the diffuse emission of the X-ray PWN, its radio counterpart was not detected in the 1-2~GHz band with an rms noise of 0.24~mJy~beam$^{-1}$.

We also reported the analysis of unpublished archival X-ray observations performed with {\XMM} towards PSR~J1826$-$1256. The new image revealed an elongated nebula, brighter around the pulsar with the emission in  the direction of the peak of the TeV source HESS~J1826$-$130. The spectral analysis demonstrated a non-thermal origin for the X-ray emission and a photon index $\Gamma$ softening with the distance to the pulsar. This study shows that the most plausible origin for HESS~J1826$-$130 is due to the IC mechanism within the PWN powered by PSR~J1826$-$1256.

\begin{acknowledgements}
The authors wish to thank the anonymous referee since his/her comments greatly improved our manuscript. 
This work is based on observations done with the {\XMM}, an ESA science mission with instruments and contributions directly funded by ESA Member States and the US (NASA). 
This research was partially funded by Argentina Grants awarded by ANPCYT (PICT 0571/11) and University of Buenos Aires (UBACYT 20020150100098BA). 
G. C, E. G. and A. P. are Members of the Carrera del Investigador Cient{\'\i}fico of CONICET, Argentina. L. D. and L. S. are fellows of CONICET, Argentina.
\end{acknowledgements}

\bibliographystyle{aa}
\bibliography{paper.bib}

\end{document}